# Deep *Chandra* Observations of ESO 428-G014: III. High Resolution Spectral Imaging of the Ionization Cone and Radio Jet Region

G. Fabbiano[a], A. Paggi[a, b,c,d], M. Karovska[a], M. Elvis[a], , W. P. Maksym[a]
and Junfeng Wang[e]

*a. Center for Astrophysics, Harvard & Smithsonian, 60 Garden St. Cambridge MA 02138, USA*

*b. Dipartimento di Fisica, Universita' degli Studi di Torino, via Pietro Giuria 1, I-10125 Torino, Italy*

*c. Istituto Nazionale di Fisica Nucleare, Sezione di Torino, via Pietro Giuria 1, 10125 Torino, Italy*

*d. INAF-Osservatorio Astrofisico di Torino, via Osservatorio 20, 10025 Pino Torinese, Italy*

*e. Department of Astronomy and Institute of Theoretical Physics and Astrophysics, Xiamen University, Xiamen, 361005, China*

Abstract

We have analyzed the deep *Chandra* observation (~155 ks) of the Compton Thick Active Galactic Nucleus (CT AGN) ESO 428-G014, to study in detail the morphology of the diffuse X-ray emission in the inner ~500 pc radius region. Comparing different X-ray energy bands we find localized differences in the absorbing column and in the emission processes. Collisional ionization may be prevalent in the area of most intense optical line emission (H$\alpha$ and [OIII]). There is a good correspondence between optical line, radio continuum and soft (<3 keV) X-ray features, consistent with simulations of jet/molecular disk interaction. At all energies >3 keV, the extended emission in the central 1''.5 (170 pc) radius circumnuclear region amounts to ~70-30% of the contribution of a point source in that area (or ~40-25% of the total counts in the region). Within a 5'' radius, the contribution from extended emission overcomes that from a nuclear point source in the 3-4 keV band. This extended emission suggests scattering of nuclear photons by dense molecular clouds in the inner galactic disk of ESO 428-G014. Its presence may adversely bias the torus modeling of spectra from X-ray telescopes with inferior angular resolution than *Chandra*, such as *NuSTAR* and *XMM-Newton*.

## 1. Introduction

This is the third paper on the results of our deep *Chandra* observations of Compton Thick (CT) AGN in ESO428-G014 (also called IRAS01745-2914, MCG-05-18-002), a southern barred early-type spiral galaxy [SAB(r)], at a distance of ~23.3 Mpc (NED; scale=112 pc/arcsec). This galaxy has a highly obscured Compton Thick (CT, $N_H>10^{25}$ cm$^{-2}$) Seyfert Type 2 nucleus, with a high ratio of [OIII] $\lambda$5007 to hard 2-10 keV observed flux (Maiolino et al 1998), and the second highest [OIII] flux among CT AGNs after NGC1068 (Maiolino & Riecke 1995, Risaliti et al 1999), so it is expected to be the second intrinsically brightest CT AGN below 10 keV.

The nucleus of ESO428-G014 is associated with a relatively weak radio source (18.9 mJy at 8.4 GHz , Leipski et al 2006; 81.7mJy at 1.4 GHz, Condon et al 1998). This source is a curved 5'' radio jet at 6 cm. (Ulvestad and Wilson 1989). The radio emission is spatially coincident with the extended H$\alpha$ and [OIII] optical line emission; a 2cm. high-resolution map of the nuclear region is also available (Falcke et al 1996, Falcke, Wilson & Simpson 1998). Based on a distinctive double loop in the line emission in the nuclear region, which is also the most intense region of radio emission, Falcke et al (1996) suggest a double helix structure, in the north side of the jet, reminiscent of that seen in NGC 4258 (Cecil et al 1992). The southern side of the jet is instead coincident with a higher excitation extended area, based on the relative excess of [OIII] emission. The biconical shape of this high excitation region led Falcke et al (1996) to conclude that it is the nuclear biconical outflow with 1400 km s$^{-1}$ winds suggested by the spectra of Wilson & Baldwin (1989). Near-IR IFU observations show general agreement between emission lines and radio morphology; they also suggest a disk perturbed by an outflow (Riffel et al 2006).

In Paper I of this series (Fabbiano et al 2017), we have reported the discovery of kiloparsec-scale extended components in the 3-6 keV hard continuum and Fe K$\alpha$ 6.4 keV line emission. These hard emission components are not confined to the nuclear surroundings, as would be expected in the model of an AGN enshrouded by a Compton thick obscuring torus (e.g., Marinucci et al 2012). Their extent suggests scattering in the interstellar medium (ISM) of ESO428-G014 of photons escaping the nuclear region in the direction of the ionization cone.

In Paper II (Fabbiano et al 2018), we have reported a detailed spectral and spatial study of the diffuse large-scale emission in the entire *Chandra* energy range (0.3-8.0 keV). The entire spectrum is best fitted with a mix of photoionizaton and thermal components, indicating the presence of varied physical conditions in the emitting plasma. The spatial extent of the kiloparsec scale diffuse emission is larger at the lower energies, suggesting that the optically thick molecular clouds responsible for the scattering of the higher energy photons are more concentrated in the inner radii of ESO 428-G014. Diffuse emission is also detected in the region perpendicular to the ionization cone, where the AGN obscuring torus in the standard AGN model (e.g. Netzer 2015) should obscure the nucleus entirely, indicating that the torus is likely to be porous.

In this paper, after reviewing the data analysis procedures (Section 2), we study in detail the inner ~500 pc radius high-surface brightness emission (Sections 3 and 4); we compare the X-ray images, both broad-band and in selected X-ray emission lines with optical emission line and radio continuum maps (Falcke et al 1996, 1998; Sections 5 and 6); and we estimate the contributions of the point-like AGN source and extended emission in the inner 170pc (1.5") circumnuclear region (Section 7). We discuss implications of our results in Section 8. We summarize results and conclusions in Section 9.

## 2. Data Reduction & Analysis

Table 1. summarizes the observations used in this paper. The data and basic reduction and analysis procedures used here follow closely those used in Papers I, and II, and we refer to those papers for a detailed description. We repeat here the basic procedures, and add specific information pertinent to this paper.

Table 1. Observation Log

| ObsID | Instrument | $T_{exp}$(ks) | PI | Date |
|---|---|---|---|---|
| 4866 | ACIS-S | 29.78 | Levenson | 2003-12-26 |
| 17030 | ACIS-S | 43.58 | Fabbiano | 2016-01-13 |
| 18745 | ACIS-S | 81.16 | Fabbiano | 2016-01-23 |

As explained in Paper I, all the data sets were screened, processed to enable sub-pixel analysis and merged. For our analysis, we used *CIAO* 4.8 tools (http://cxc.cfa.harvard.edu/ciao/) and the display application *DS9* (http://ds9.si.edu/site/Home.html), which is also packaged with *CIAO*. To highlight some of the features, we applied different Gaussian kernel smoothing to the images, including both 1-kernel Gaussian smoothing and multi-kernel adaptive smoothing (with *Asmooth* in the DS9 CIAO package), as described in the text and figure captions (Section 3). The *Chandra* PSF was simulated using rays produced by the *Chandra Ray Tracer* (*ChaRT*) projected on the image plane by *MARX* (*http://space.mit.edu/CXC/MARX/*). As in Paper II, we have also applied the image restoration algorithm Expectation through Markov Chain Monte Carlo (EMC2, Esch et al. 2004; Karovska et al. 2005, 2007; Wang et al. 2014; sometimes referred to as PSF deconvolution) with the appropriate PSF model (Section 4).

We have used subpixel binning of the ACIS data to produce our images, trying to go for the highest resolution needed to highlight features in the data. The ACIS instrument pixel is 0''.492 on the side, so that the point response function (PSF) is under-sampled. As already discussed in Paper II, higher effective resolution can be achieved by binning the data in smaller pixels to oversample the PSF (subpixel binning). This method was used in and validated by our previous work on AGN circumnuclear regions, where we have demonstrated how the features recovered from the ACIS images match well with both higher instrument resolution *Chandra* HRC images, optical emission line clouds resolved with *HST*, and high-resolution radio features (e.g. Wang et al 2011a, b, c; Paggi et al 2012).

To characterize features emerging from the maps, we have in all cases extracted the integrated counts in regions from the raw unsmoothed data, and explained this procedure in the relevant text.

Figures 1 and 2 show both 1/16 (0''.031 pixel) and 1/8 (0''.062 pixel) subpixel imaging of the inner bright regions of the diffuse emission, leading to our choice of 1/16 subpixel binning for most of the maps shown in this paper. Fig. 1 shows (images on the left) the raw binned ACIS data (1/16 on the top and 1/8 on the bottom), with the resulting adaptively smoothed images on the right. For both, 5 counts under the kernel were used, with Gaussian kernels ranging from 2 to 15 image pixel, in 30 steps. While the overall appearance of the diffuse emission is similar, important details are smoothed out in the 1/8 subpixel image. These are real significant features (see Fig. 2 for a zoomed-in image), as can be seen from extracting counts from the regions highlighted in the image from the raw data (upper right in Fig. 1 and right in Fig. 2; The counts are written within or nearby the relevant regions.) These features can be recovered if we perform adapting smoothing on the 1/8 subpixel data, starting from a kernel of 1 image pixel, but the resulting images shows obvious pixellation (Fig. 2, Left). For this reason, we have used the 1/16 subpixel in most of the images shown in this paper.

At the soft energies contributing most of the photons in these images, a circle of 1'' diameter will contain about $\geq 75\%$ of the encircled energy. The circular regions that we have used for count extraction have radii of 0''.15, which empirically we consider the resolution of these images. This is also empirically confirmed by comparison with high-resolution optical line and radio continuum images (Sections 3-5). However, 0''.15 radius circle contains ~0.3 of the encircled energy of the *Chandra* PSF. Because of the shape of the PSF, each of these extraction regions will also contains counts from the adjacent regions. Therefore, the 'at the source' differences in signals are attenuated, and the physical differences between the features are likely to be larger than we can report.

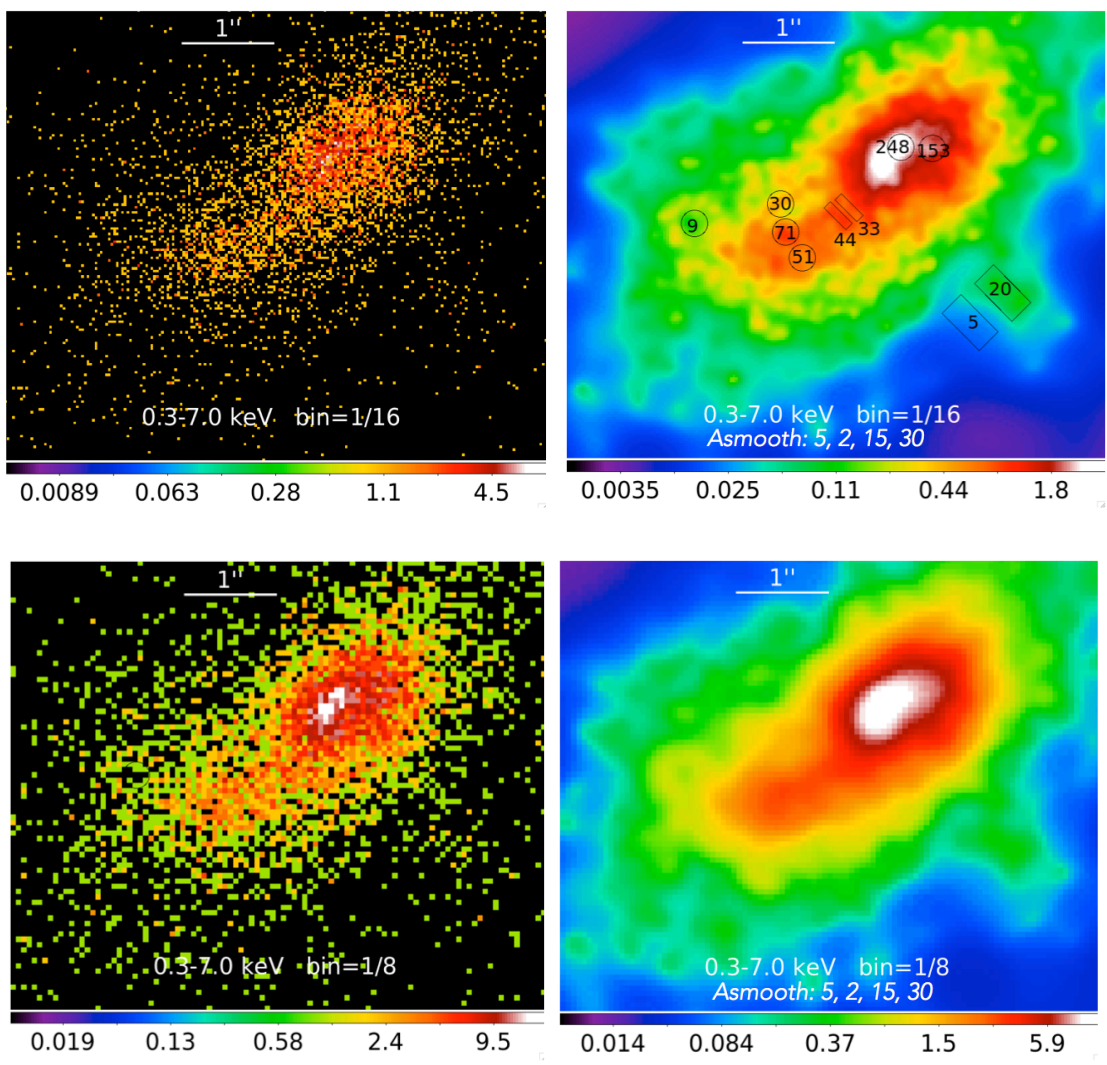

Fig. 1 – Top: 1/16 subpixel raw and adaptively smoothed data as indicated on the image. Bottom: the same for 1/8 subpixel data. Counts extracted from the raw data from identifiable features in the 1/16 adaptively smoothed image are shown in the top-right image with the extraction regions. Most of these features are not identifiable from the 1/8 subpixel image in the bottom right. The intensity scale is logarithmic and the color scale is in counts per image pixel for all the images.

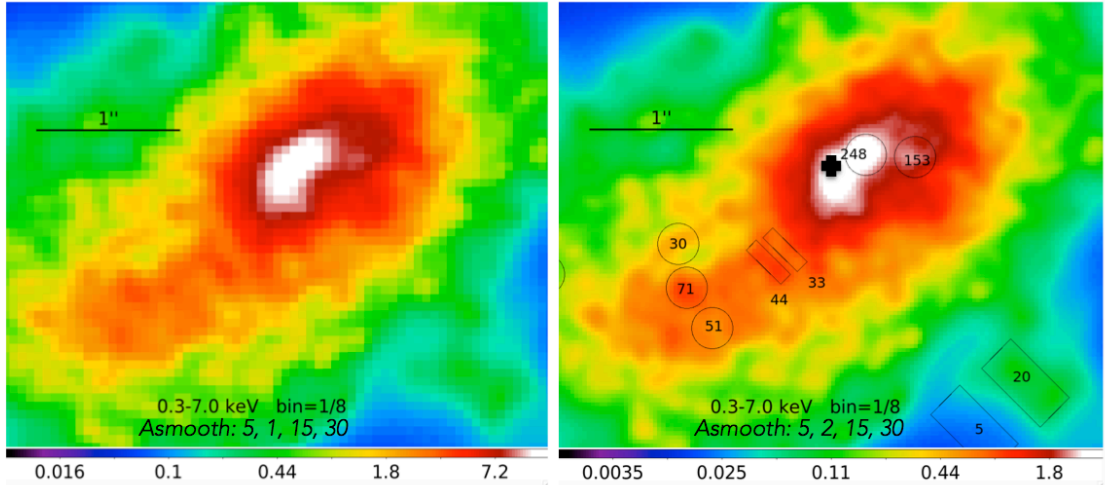

Fig. 2. Comparison of 1/8 subpixel image with adaptive smoothing starting from 1 image pixel versus the 1/16 subpixel image with smoothing starting from 2 image pixels (same as in Fig. 1 top-right). The variations in surface brightness are similar, but the pixel boxiness is visible in the 1/8 pixel image on the left, when compared to the 1/16 pixel on the right. The intensity scale is logarithmic and the color scale is in counts per image pixel for each image. The black cross is at the centroid of the hard (4-6 keV) nuclear source (J2000 RA = 7:16:31.21, Dec = -29:19:28.6; See Fig. 7).

## 3. The Central ~500 pc High X-ray Surface Brightness Emission

Fig. 3 shows the inner high surface brightness region of ESO 428-G014 in the soft emission-line dominated 0.3-3.0 keV band and in the hard continuum emission 3.0-6.0 keV band (see Paper II for the spectral characterization of the emission). Fig. 4 shows a finer subdivision of the soft emission into 0.3-1.5 keV and 1.5-3.0 keV bands. The data were binned with 1/16 pixel to exploit the full resolution in the high-count inner regions, and processed with Gaussian adaptive smoothing (*Asmooth*: 2-15 pix scales, 5 counts under kernel, 30 steps). Contours have been added to highlight the main features of the 0.3-3.0 keV image. The same contours are also plotted on the other images, for ease of comparison.

Overall, the X-ray emitting region has a characteristic head-tail morphology, where the head comprises the 1'' (112 pc) radius circumnuclear region, and the tail extends to the SE for 2-4'', depending on the energy band. However, there are some clear energy-dependent differences in the morphology of the X-ray surface brightness. As can be seen in Fig. 3 (bottom), the nuclear source has a definite peak in the 3.0-6.0 keV image, although embedded in a region of diffuse high surface brightness (see the comparison with the PSF in Fig. 5). Instead, the central higher surface brightness pixels in the 0.3-3.0 keV image (Fig. 3, top) are distributed along the same axis as the 'tail' extended component. In the 0.3-1.5 keV image (Fig. 4, top), the nuclear source is not visible, consistent with the high CT column density in the nuclear regions.

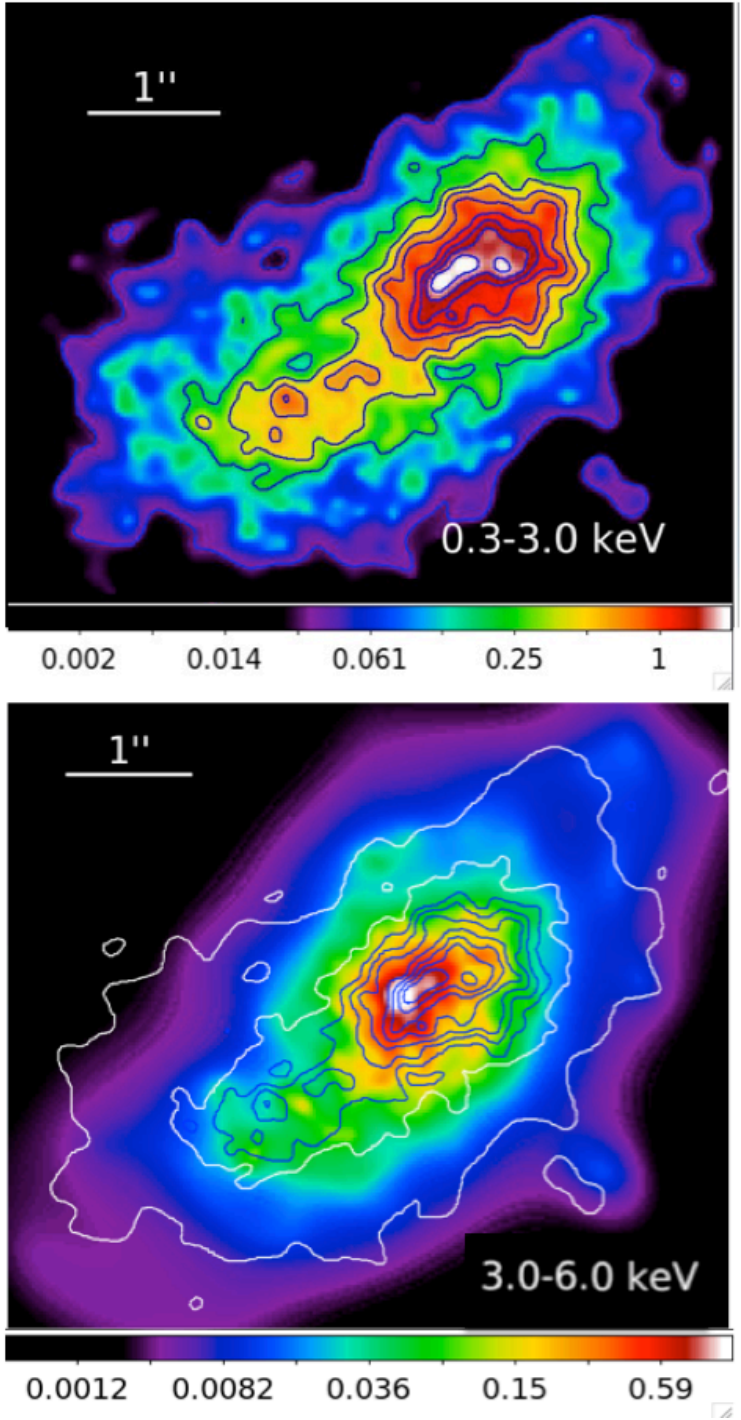

Fig. 3 – Adaptively smoothed images in the indicated energy bands (1/16 subpixel binning; 2-15 pix scales, 5 counts under kernel, 30 steps). The contours are derived from the 0.3-3.0 keV image and plotted on the other image to facilitate comparison of features. The image intensity scale is logarithmic. The color scale is in counts per 1/16 ACIS pixel. The peak of the X-ray emission in the 3.0-6.0 keV image is at the hard nucleus.

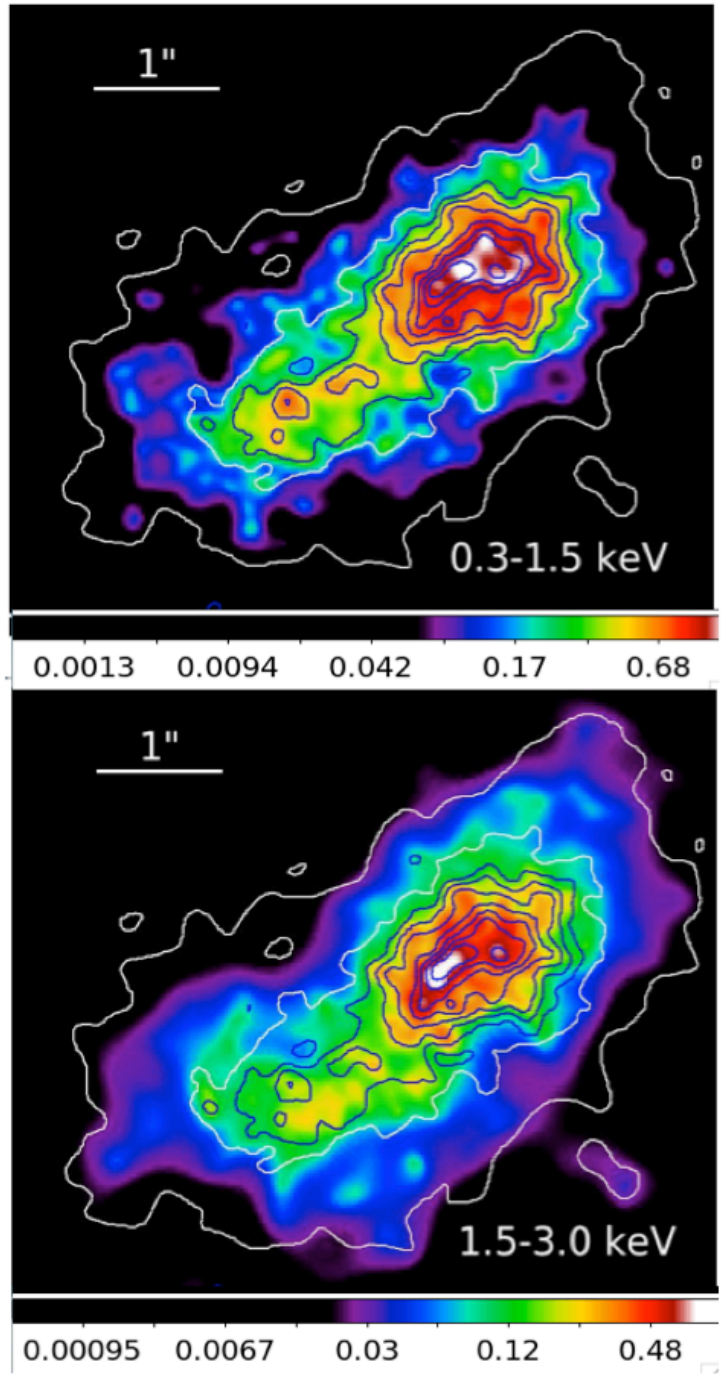

Fig. 4 – Adaptively smoothed images in the indicated energy bands (1/16 subpixel binning; 2-15 pix scales, 5 counts under kernel, 30 steps). The contours are derived from the 0.3-3.0 keV image (Fig. 3) and plotted on the other images to facilitate comparison of features. The image intensity scale is logarithmic. The color scale is in counts per 1/16 ACIS pixel.

As discussed in Paper II, the diffuse emission is more prominent at the lower energies. However, even at the higher energies >3.0 keV, where there is a clear single peak of surface brightness (Fig. 3, bottom), the emission of the nuclear region cannot be explained with a single point source. This is shown clearly by the radial profiles of Fig. 5. Following the procedures of Paper II, we have produced these azimuthally averaged radial profiles centered on the highest surface brightness pixel of the 3-6 keV image (the nuclear source); the PSF was normalized to this central surface brightness. Note that these radial profiles extend past the region shown in Fig. 3. The extended emission at large radii, including the azimuthal dependencies, are fully analyzed and discussed in Paper II. In Section 7 we will revisit the budget of extended versus point-like emission in the circumnuclear region at energies above 3.0 keV.

We also find localized differences in the 0.3-1.5 keV and 1.5–3.0 keV images, which we will explore below. To highlight the differences between softer and harder emission images, we have created ratio maps of selected energy bands, binned at ¼ pixel resolution (0''.125 pixel), so as to enhance counting statistics. Fig. 6 shows the (3.0-6.0 keV)/(0.3-3.0 keV) ratio image on the top and the (0.3-3.0 keV/0.3-1.5 keV) on the bottom, with 2 image pixel Gaussian smoothing. Both ratio images show a band of high value pixels (harder emission), extending NE to SW across the nuclear region. Average ratios with 1σ errors (in parentheses) are given in the right panels of Fig. 6, for the indicated areas, demonstrating that the differences highlighted by the ratio image are significant. Hard excess is visible at the nuclear position, not surprising since the flat continuum 'emerges' above 3 keV (Paper II). There also excesses flanking the nucleus in the direction perpendicular the main axis of the soft emission extent (the 'cross-cone' direction of Paper II). These are probably due to the markedly elongated geometry of the soft emission, possibly connected with the presence of a radio jet (see Section 6.), when compared to the rounder hard distribution (Fig. 3).

To highlight differences below 3.0 keV, we chose to plot the (0.3-3.0 keV/0.3-1.5 keV) ratio. The advantage of this ratio is that it maximizes the count statistics in each band; larger departures from unity will indicate regions of more prominent 1.5-3.0 keV emission. We find regions of high values (i.e. a relative excess of 1.5-3.0 keV emission) to the north of the head and south of the tail, respectively. The pixel count in the external regions is lower, typically <10, so counting statistics is a concern. However, the fact that these high ratio points are correlated suggests a real difference between the 0.3-1.5 keV and 1.5-3.0 keV emission in these regions, as indicated by the average values in the highlighted regions. These localized excesses are also suggested by comparing the band images in Fig. 4. The high 0.3-3.0 keV/0.3-1.5 keV ratio region that we find to the north of the head appears to follow the edge of the region with larger J-K colors of Riffel et al (2006), suggesting a connection with absorption of the softer photons in these regions. For a steep spectrum, such as that below 3 keV (Paper II), the ratios may be indicative of absorbing clouds with $N_H \sim$ several $10^{21}$ –a few $10^{22}$ cm$^{-2}$.

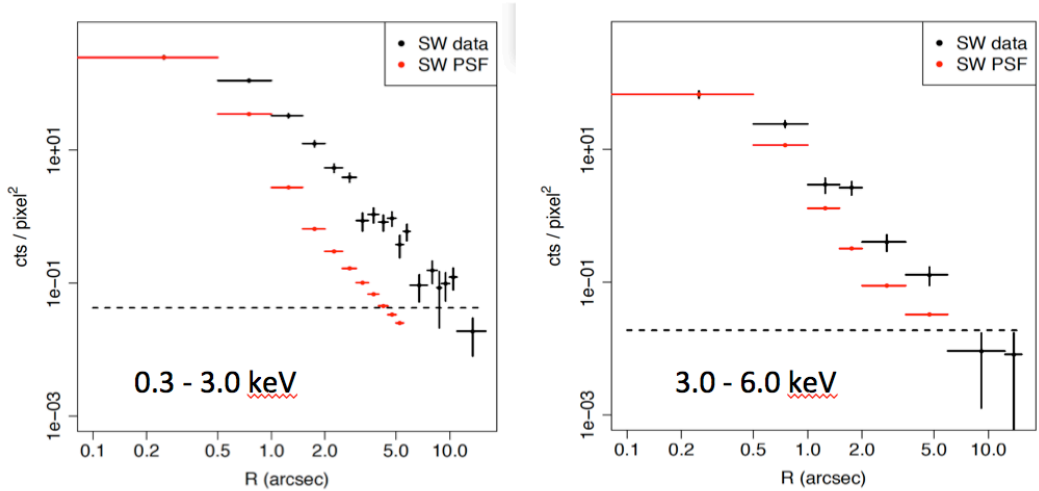

fig. 5 – The radial distribution of the X-ray surface brightness to the SW of the nucleus in two energy bands (see Paper II). Both profiles were centered on the centroid of the 3-6 keV nuclear source (see Fig. 3, bottom) and the PSF was normalized to this surface brightness.

### 3.1. X-ray emission line maps

We have further explored the localized differences in the extended soft X-ray surface brightness, at different energies (Fig. 4) by means of narrow-band images, encompassing prominent X-ray emission lines. Given that the emission below 3 keV is dominated by an ionized medium (Paper II), some of these differences, if not due to absorption, are likely to be related to the localized physical state of this medium.

Fig. 7 shows three narrow-band images, containing the O VII (0.40-0.62 keV), Ne IX + Ne X (0.9-1.2 keV), and Si XIII (1.65-2.0 keV) doublet emission lines (see Paper II). For comparison, we also show the 4-6 keV image, dominated by the nuclear source. Overlaid on all the images are the Hα contours from the images of Falcke et al (1996; see Section 5. below).

All the line images in Fig. 7 show extended emission that follows the Hα extension to the SE, but with differences. The peak of the O VII emission (Fig. 7 top left) is consistent with the position of the nuclear source in the hard, 4-6 keV, continuum (Fig. 7 bottom left). A secondary peak occurs at the position of the bend of the Hα contours to the SE. The brightest regions of Ne IX, X emission (Fig. 7 top right) instead coincide with the northern Hα loop. The Ne IX, X tail is also pretty much coincident with the Hα tail, except for a possible ridge of Ne IX, X to the north of this tail. The Si XIII emission in the nuclear region (Fig. 7 bottom right) appears to be spatially complementary to the Ne IX, X.

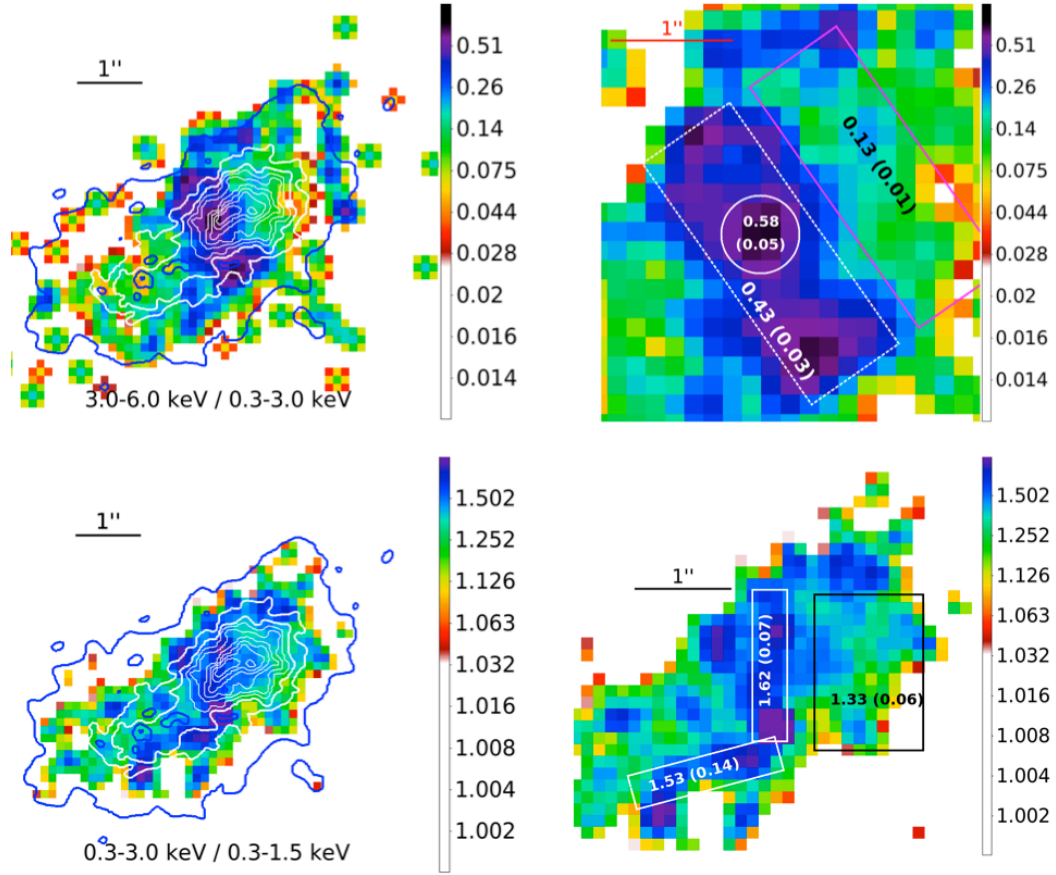

Fig. 6 – Top: left - ratio of the 3.0-6.0 keV and 0.3-3.0 keV images; right – a zoom in of the same area with average ratio values (1σ error in parentheses) in the marked regions. Bottom: left – ratio of the 0.3-3.0 keV and 0.3-1.5 keV images; right - a zoom in of the same area with average ratio values (1σ error in parentheses) in the marked regions. These images are derived from ¼ subpixel data with 2 pixel Gaussian smoothing. The 0.3-3.0 keV contours of Fig. 3 are superimposed for reference. The image intensity scale is logarithmic. The color scale is in ratio value per 1/4 ACIS pixel.

To explore the implications of these differences, we have extracted counts from the three high-surface brightness, not overlapping spatial regions, easily identifiable in Fig. 7, centered on the O VII nucleus and SE knots, and in the region of Ne IX+X excess NW of the nucleus. In all cases we have used a circular region with 0".3 radius. These regions contain the central peak of the PSF. However, because of the PSF shape they will also be somewhat contaminated by counts from the surrounding areas, which may have some attenuating effect on spectral differences.

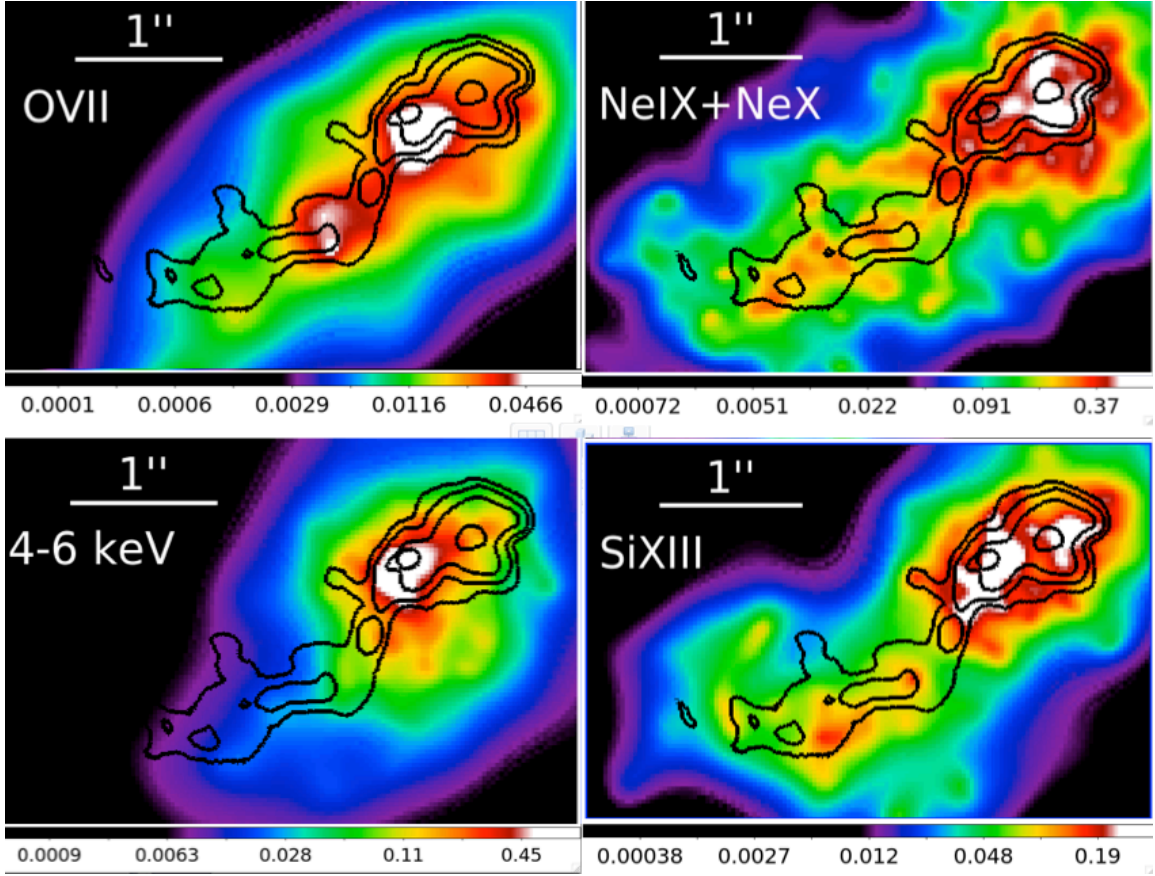

Fig. 7 - Images in energy bands corresponding to the line emission indicated in the figure. 1/16 pixel, with adaptive smoothing (5 counts under Gaussian kernel, scales from 2 to 15 pixel, 30 steps). From top left, clockwise: 0.40-0.62 keV, 0.9-1.2 keV, 1.65-2.0 keV, 4-6 keV continuum. High contrast color scales were used to emphasize the high surface brightness features. H$\alpha$ contours superimposed. The image intensity scale is logarithmic. The color scale is in counts per 1/16 ACIS pixel. The counts to calculate the line ratios were extracted from 0".3 radius circles centered on the high surface brightness ('white') regions in the figures: the nucleus (as visible in the 4-6 keV image, and also OVII), the SE OVII knot at the bend of the H$\alpha$ contours, and the intense NeIX + NeX area NW of the nucleus.

We have extracted the counts from each of these areas and calculated the line ratios, using bandwidths of 0.40-0.62 keV, 0.90-1.20 keV and 1.65-2.0 kev for the O, Ne, and Si lines respectively. These ratios were then compared to those estimated from simulations using a single component photoionization CLOUDY model (Ferland et al 1998), and the instrumental parameters appropriate for our observations. From these models we simulated a grid of expected line ratios for a range of model parameters consistent with the spectral analysis of Paper II (log U$\sim$−3 to 2, log $N_H \sim$ 19 to 22.5). Fig. 14 (see Section 8) shows the position of the measured ratios in the 3-D space of simulated photoionization ratios, for different values of log U and log $N_H$. All three measured ratios have different emission parameters, taking into account their statistical errors.

## 4. The Soft X-ray Emission (0.3-3.0 keV) from the Central ~200 pc

If we ignore the energy-dependent variations of the spatial emission (Fig. 7), we can attempt to explore the overall morphology of the ionized gas in the soft band (0.3-3.0 keV) with EMC2, and good statistical significance. In this energy band there are over 2400 counts within the 1'' circle (~112 pc radius). Fig. 8 (left) shows the data binned with 1/16 pixel and smoothed with a 4 pixel Gaussian. The black circle in the figure marks the position of the nuclear source in the 4-6 keV continuum (see also Paper IV, Fabbiano et al 2018c). We expect, based on a MC simulation of a source with the observed ~170 counts (4-6 keV), the 1σ statistical uncertainty in the centroid position to be ~0.1''. The areas of most intense emission in the 0.3-3.0 keV band (Fig. 8 left) show a linear jet-like structure ~50pc in length extending to the NW from the nucleus, ending up in a bubble-like structure.

On the right, Fig. 8 shows the EMC2 PSF restoration of the same area (which corrects for the effects of the PSF spread), resulting from the average of 100 iterations. We caution that a full calibration of the core of the PSF is still pending, so that some unknown systematics may affect our models. We will explicitly address these issues in our analysis of the inner nuclear region (Paper IV).

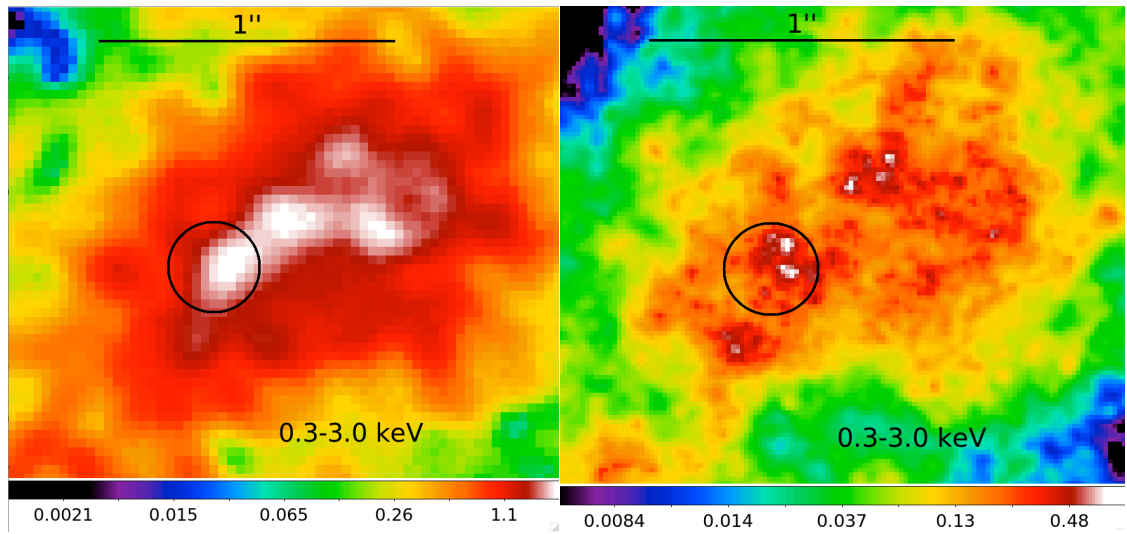

Fig. 8 – The circumnuclear region in the 0.3-3.0 keV band. Logarithmic scale. Subpixel binning of 1/16. Left, smoothing with a Gaussian kernel of 4 image pixels; right, EMC2 restoration and 0.5 -1 pixel Gaussian adaptive smoothing, to enhance the filamentary structure. The black circle marks the position of the nuclear source in the 4-6 keV continuum in the raw (see Fig. 7, bottom left) image. The image intensity scale is logarithmic. The color scale is in counts per 1/16 ACIS pixel.

The 0.3-3.0 keV EMC2 restored image suggests a frothy bubble structure with two regions of enhanced emission, one at the nuclear position, and the other to the NW, at the base of two loop-like filaments. Comparison with Fig. 7 shows that the emission of the bright knot may be dominated by the Ne IX, X and Si XIII lines.

## 5. Comparison with Optical line and Radio emission

*Chandra* imaging studies of Seyfert galaxies (e.g., Bianchi et al 2006; Wang et al. 2011a, b, c; Paggi et al. 2012) have generally demonstrated a good correspondence between optical line and soft X-ray extended features. We therefore used the soft emission-line dominated 0.3-3.0 keV band to compare the X-ray, H$\alpha$, [OIII] emissions. To compare more directly the soft X-ray and H$\alpha$ images, we have generated contours highlighting the higher surface brightness features of the H$\alpha$ emission (left panel in Fig. 9; from Falcke et al 1996), and overlaid them onto the 0.3-3.0 keV image.

The H$\alpha$ image coordinates were converted to J2000, to match the *Chandra* coordinates. After this correction, we shifted the contours to the NW by 0.4'', consistent with the overall astrometrical uncertainties of the two data sets, to match the position of the optical nucleus (Falcke et al 1996, the cross in the H$\alpha$ image, Fig. 9 left) with the centroid of the 4-6 keV hard X-ray nuclear source (the blue circles in Fig. 9). Fig. 9 (top, right) shows a high resolution (1/16 pixel) high contrast image of the 0.3-3.0 keV emission, compared with the H$\alpha$ *HST* image (Falcke et al 1996). Fig. 9 (bottom, right) shows the shifted H$\alpha$ contours over the 0.3-3.0 keV image obtained by applying the EMC2 image restoration (averaging 100 iterations). The resulting correspondence between H$\alpha$ and X-ray features is remarkable, especially with the EMC2 reconstructed image, both in the head and in the tail to the SE. We used the same shift to match the [OIII] contours, which are similar to the H$\alpha$, although less well defined. The areas of stronger X-ray emission are spatially coincident with the 'crossing' in the H$\alpha$ strands. These are also the area of most intense localized [OIII] emission (Fig. 10).

Fig. 10 shows the overlay of the H$\alpha$ and [OIII] contours (Falcke et al 1996), on the larger scale 0.3-3.0 keV X-ray emission. We show here the EMC2 restored images, derived from 1/8 pixel binning, so that to have enough statistics to cover the outer low-surface-brightness areas. In this image, the second crossover of the H$\alpha$ strands, NW of the nucleus, appears as the highest surface brightness spot. The overall agreement between optical line contours and X-ray brightness levels (from the color scale) is remarkable, except for the SE corner, where the [OIII] emission stretches further out (as noted by Falcke et al 1996, see Fig. 10 bottom left, showing the [OIII]/H$\alpha$ contours). Fig. 10 (bottom right) shows the overlay of the 6 cm. contours on the same soft X-ray 0.3-3.0 keV image. The upturn of the radio jet (Fig. 10 bottom right) occurs just past the region of intense H$\alpha$ and X-ray emission shown in Fig. 9, in proximity of the upturn in the Si XIII band emission (Fig. 7).

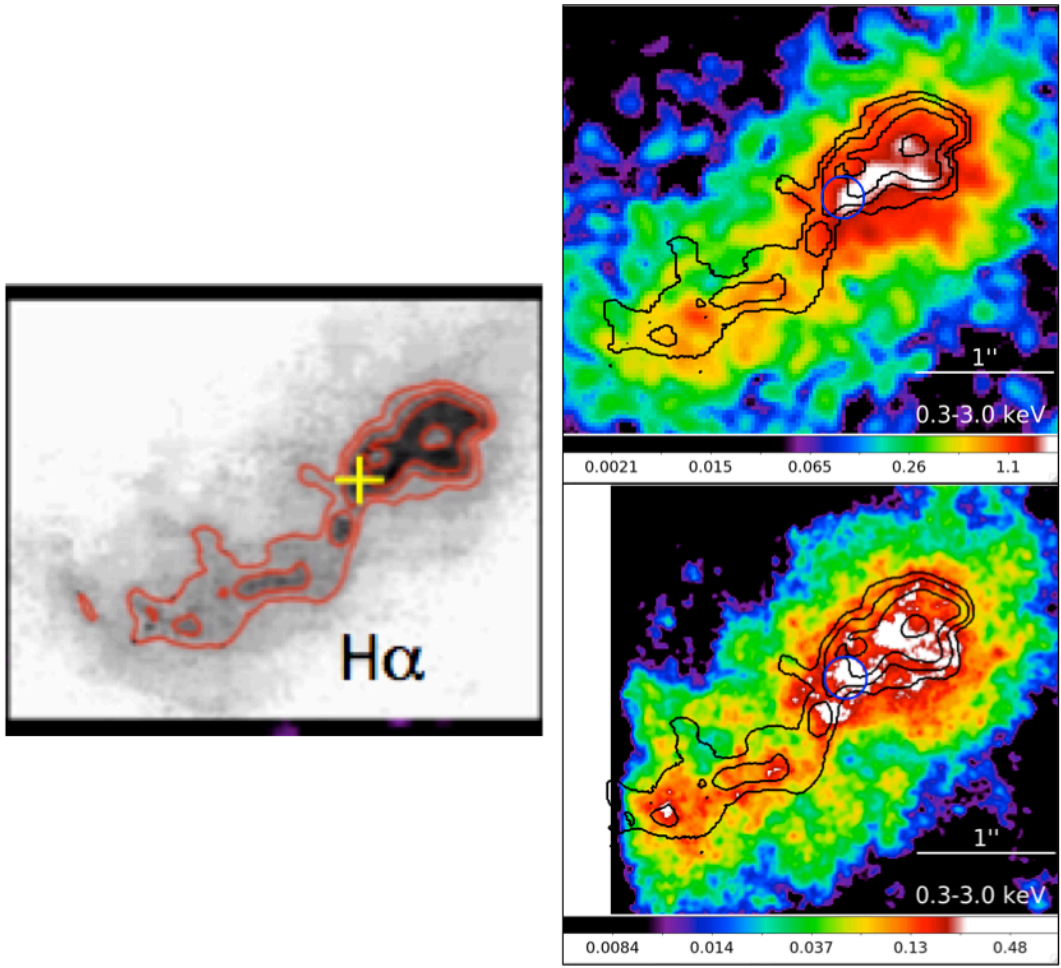

Fig. 9 – Left: Hα image and derived brightness contours. The cross marks the position of the nucleus (Falke et al 1996). Top: 0.3-3.0 keV 1/16 ACIS pixel image smoothed with a 4 image pixel Gaussian. Bottom: 0.3-3.0 keV 1/16 pixel EMC2 image restoration, with enhanced contrast to show the highest surface brightness areas. The image intensity scale is logarithmic. The color scale is in counts per 1/16 ACIS pixel. The contours are from the Hα image (see text). The circle is at the position of the 4.0-6.0 keV nuclear source.

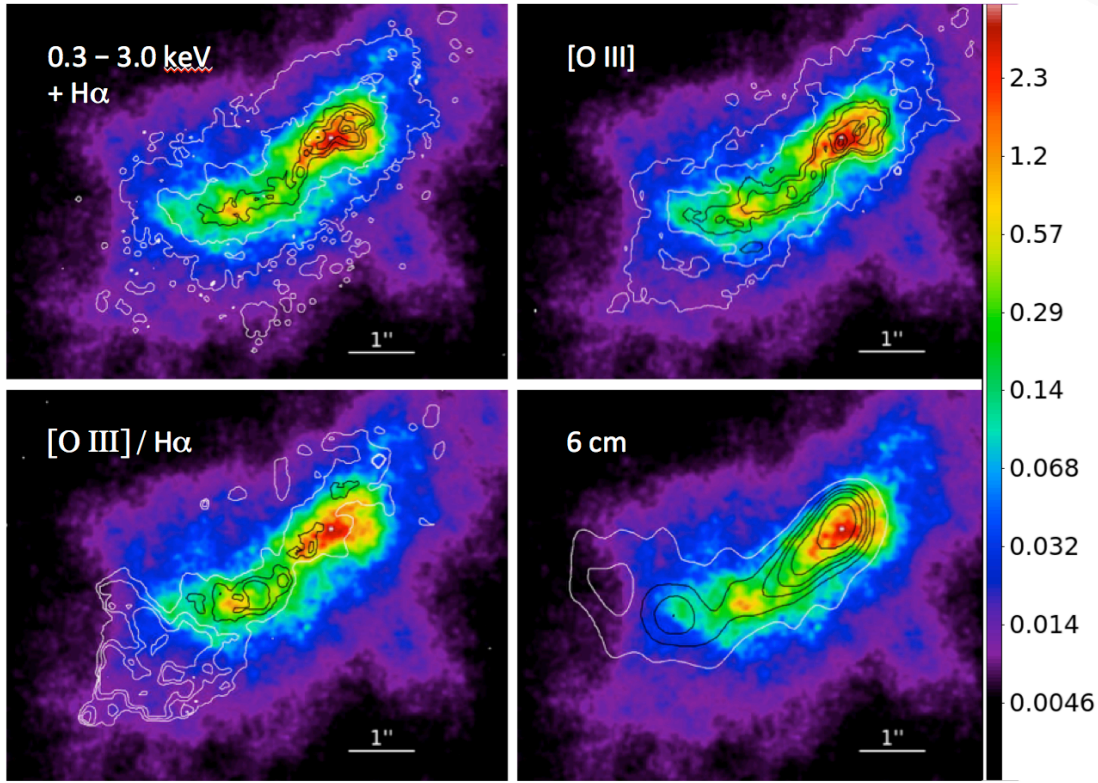

Fig. 10 - 0.3-3.0 keV 1/8 subpixel EMC2 restored image with 2 pixel Gaussian smoothing. Top: Hα (left) and [OIII] (right) contours. Bottom: [OIII]/Hα (left) and 6cm. (right) contours (from Falke et al 1996). Contours are linear and cover the entire range of the respective image. The image intensity scale is logarithmic. The color scale is in counts per 1/8 ACIS pixel.

ESO428-G014 has also been studied by Riffel et al. (2006) in near-infrared emission lines using the Gemini GNIRS IFU in natural seeing (~0.8" – 0.9" FWHM). $H_2$ (2.121 μm), Br γ, Pa β, and [FeII] (1.257 μm) images were obtained. Overall the morphology of these lines follows that of the 2 cm radio image (Falcke et al. 1988) and the X-rays in the inner 2 arcsecond radius region. In more detail there are differences; the [FeII] line is weaker in the SE and extends further to the south in the NE than the $H_2$ line. This [FeII] morphology may correspond better to the soft X-ray map. [FeII] is created by lower ionization photons than $H_2$. A more detailed comparison with the X-ray emission lines is hindered by the lower angular resolution of the GNIRS data and the limited dynamic range of the *Chandra* data.

## 6. Comparisons between X-ray, Radio Continuum and Hα Features in the Inner Nuclear Region

Falcke et al (1998) presented a 2 cm high resolution VLA image of the nuclear source, and compared it with the Hα image, to advance a picture of the nuclear emission of ESO 428-G014. Based on this comparison, they identified the radio nucleus with the SE bright knot in the radio image (see contours in Fig. 11) and by matching 2 cm and Hα contours so that the eastern bright 2 cm knot overlaps with the first crossing of the Hα strands (see cross in the insert in Fig. 9), they concluded that the radio emission hugs the northern Hα strand. The first crossing of the Hα strands corresponds with the nuclear position, based on the centroid of the optical stellar continuum emission (Falcke et al 1996, the cross in the insert of Fig. 9), which we have matched to the hard continuum nucleus in the X-rays (Section 5). While the match between optical nucleus, Hα and hard-X-ray nucleus is strong, we believe that the match with and identification of the radio nucleus is still somewhat ambiguous.

Fig. 11 (left) shows the 2 cm contours of Falcke et al (1998), shifted to J2000 coordinates and plotted over the hard continuum nuclear peak (smoothing only). The X-rays nucleus is near the central fainter 2 cm knot, a shift of 0.15" reconciles the two centroids.

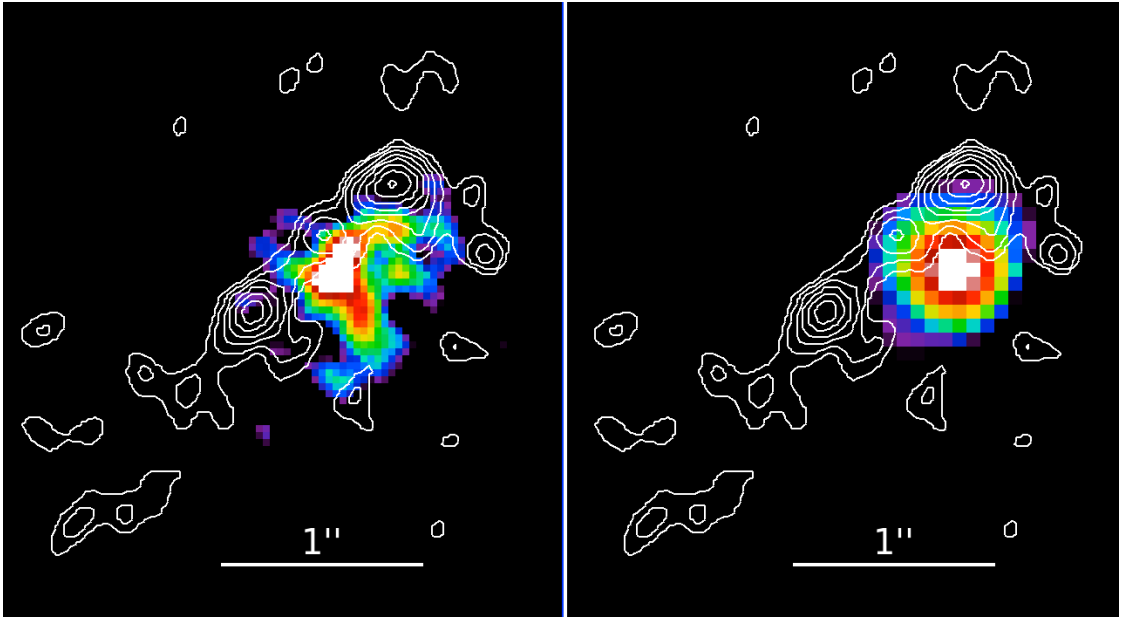

Fig. 11 – Left: high contrast 5.0-6.0 keV image of the central region, showing the prominent nuclear X-ray source. Superimposed are the 2cm VLA contours from Falcke et al (1998). The X-ray nucleus is only ~0.15" apart from the center of the central radio source. Right: EMC2 restored image in same energy band. The X-ray centroid is shifted 0.3" from the central radio knot.

Fig. 11 (right) shows the overlay of the 2cm contours on the EMC2 restoration of the same image. There is a small shift (0.2'') between the centroid in the raw data and in the restored image. Even in this case the comparison with the 2 cm contours points to the central radio knot as the likely counterpart. If this identification is correct, the radio source is not a one-sided jet, as suggested by Falcke et al (1998), but a symmetrical two-sided radio source. Such symmetrical radio sources are common in radio-quiet AGNs (Wilson and Willis 1980). However, the Falcke et al (1998) solution is certainly within the *Chandra* astrometric uncertainties.

If we assume the alternate picture that the central weaker radio knot is the nucleus, we can extend this comparison to include the Hα contours and the soft extended X-ray emission from the EMC2 restored image (Fig. 12).

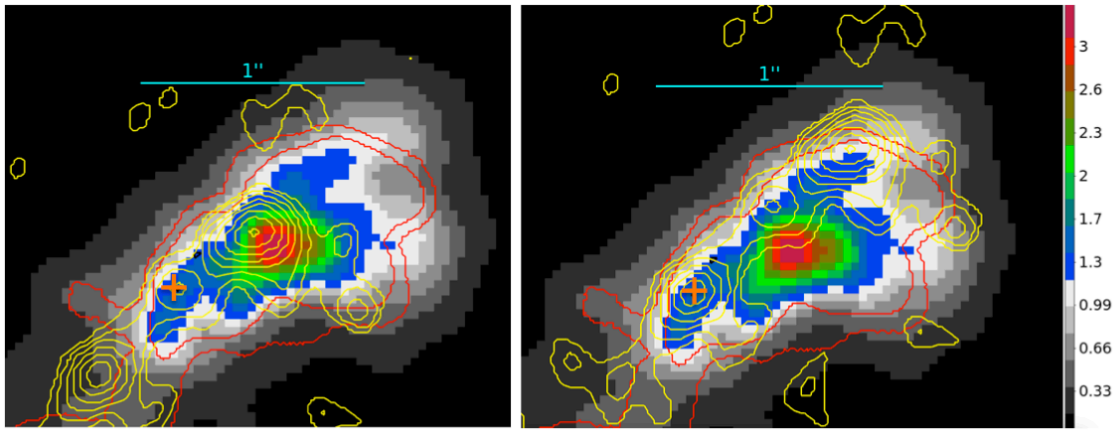

Fig. 12 - 0.3-3.0 keV 1/8 pixel EMC2 restored X-ray images with Hα (red; for visibility we have eliminated the inner contours) and 2cm (yellow) contours. Left: 2cm contours are shifted so that the central weaker radio source matches the hard nuclear source (orange cross). The orange cross is also the position of the optical nucleus and of the first crossover of the optical line emission strands (see Fig. 9 and text). With this shift, the NW bright radio knot coincides with the second crossover of the optical emission lines, with is also the region rich in Ne IX, X and Si XIII emission (Fig. 7). Right: 2cm contours shifted so that the SW knot matches the hard nuclear source (as in Falcke et al 1998). This is the same image shown in Figs. 8 and 9, but with a different color scale, to highlight the regions of highest X-ray surface brightness. The image intensity scale is logarithmic. The color scale is in counts per 1/8 ACIS pixel.

We find that the bright X-ray knot at the second crossover of the Hα and [OIII] strands (Figs. 9, 11; Falcke et al 1996), corresponding to an X-ray bright spot in the soft (<3 keV) energy band, could be associated with the bright western radio knot. This X-ray bright spot is also the position of brightest [OIII] emission (see Fig. 10, top right). In this picture, the eastern bright radio knot would be coincident with the local drop of X-ray surface brightness at the eastern end of the high X-ray surface region within 1'' from the nucleus, and not with the nucleus, contrary to Falcke et al (1998). However, Falcke et al also remark that the exact radio/optical registration of ESO 428-G014 is somewhat uncertain. As they remark, if the radio nucleus were coincident with the southern radio knot, then the radio jet would not go through the Hα 'figure of eight', as it would be expected if the Hα were the results of shocks in the boundary layer between the radio jet and the ISM. To make this happen, the southern bright radio knot would not be coincident with the nucleus. Our alternative alignment may solve this problem.

No matter how the radio contours are shifted, the radio contours in the tail to the E appear to avoid the regions of more intense X-ray emission, which instead coincide with higher Hα contours (Fig. 13).

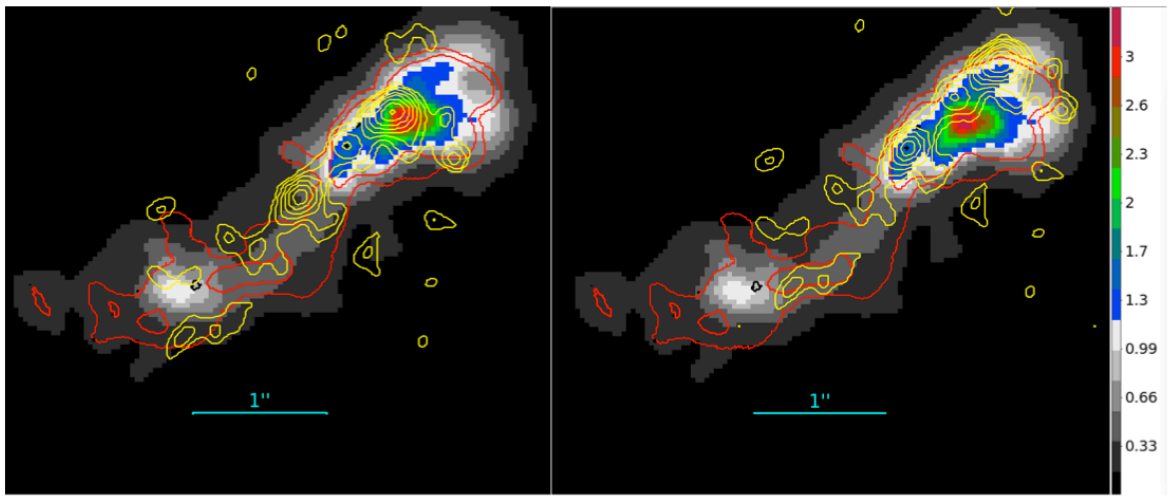

Fig. 13 – 0.3-3.0 keV 1/8 pixel EMC2 restored X-ray image with with Hα (red; for visibility we have eliminated the inner contours) and 2cm (yellow) contours. Left: the 2cm contours (yellow) are shifted so that the central knot matches the hard X-ray source. Right: 2cm contours shifted so that the SW knot matches the hard nuclear source (as in Falcke et al 1998). The image intensity scale is logarithmic. The color scale is in counts per 1/8 ACIS pixel.

## 7. The Extended Emission within the Central ~200pc at energies > 3 keV

The spectral analysis of Paper II has shown that the emission between 3 and 6 keV is a featureless hard continuum. A strong neutral Fe Kα line with the contribution of a much weaker Fe XXV line dominates the spectrum in the 6-7 keV band. In Paper I and II we have also shown how imaging in the hard continuum and the Fe Kα line revealed the presence of kiloparsec extended emission in both bands. Here we take a closer look at the hard emission in the central nuclear region.

Although an identifiable nuclear source is visible at energies above 3 keV (e.g., Figs. 3, 7), there is also a sizeable extended emission component within the inner ~100-200 pc (Fig. 5). To estimate the relative contributions of point source and extended component to the total emission of the circumnuclear region, we have extracted the counts within radii of 0".5, 1".5 and 5" from the centroid of the count distribution associated with this source, in different energy bands (see Table 2.) and compared them with the expected encircled energy for a source of characteristic energy 4.5 keV (see Jerius 2014, see also Chandra Proposers' Observatory Guide Rev. 20.0, Dec. 2017; the results for higher energy sources do not change appreciably, given the statistical errors). In all the energy bands we have assumed that the counts within the 0".5 radius are from a point source, and from this we have estimated the expected contribution from this source to the counts within larger radii.

We checked that (within statistics) possible smearing from aspect uncertainties and merging would not affect significantly the core of the PSF, by verifying the PSF encircled energy curve with nearby point sources. There are two point sources in the 3 - 7 keV band, within 25" of the nucleus (aim-point); for both, given the count statistics and the local background, the radial count distribution is consistent with the encircled energy estimates. A third point source, 87" from the nucleus instead shows a slightly more flattened PSF. To check for possible artifacts introduced by subpixel binning, we also extracted the counts with different bin sizes, including the native instrument bin, finding no difference within statistics. To be conservative, we also produced a simulated PSF (which includes a 'blur' to take into account possible aspect smearing).

Table 2 lists the measured and expected point source counts (in parentheses using the mirror PSF with no aspect blur, the square brackets identify values from the simulated PSF) within the extraction circles, the counts that cannot be attributed to a point source (Δ), and the % ratio between these and the point source counts. The errors contain 1σ counting statistics errors plus a 3% systematic error to account for the differences between mirror and simulated PSF. Field/instrumental background does not affect these estimates.

**Table 2. Extended Counts within a Radius and as Fraction of Point Source (PS)**

| Band (keV) | Counts R <0".5 | Counts [a,b] R<1".5 | Counts [a,b] R<5" | Δ [b,c] <1".5 | Δ [b,c] <5" | % Ext.[b,d] <1".5 | % Ext. [b,d] <5" |
|---|---|---|---|---|---|---|---|
| 3-4 | 128 | 297 (171 [184]) | 430 (183 [195]) | 126[113] ±18 | 237 [235]±22 | 73 [61] | 120 [120] |
| 4-5 | 123 | 260 (164 [177]) | 322 (176 [187]) | 96 [83]±17 | 146 [135]±19 | 58 [47] | 83 [72] |
| 5-6 | 134 | 236 (179 [193]) | 275 (192 [204]) | 57 [43]±16 | 83 [71]±18 | 32 [22] | 43 [35] |
| 6-7 | 200 | 374 (267 [288]) | 437 (286 [304]) | 107 [86]±21 | 151 [133]±23 | 40 [37] | 53 [44] |

a) Image counts within the indicated radius. The values in parenthesis are the estimated PSF counts.
b) Values in square brackets are from the simulated PSF, which contains aspect blur.
c) Difference between image counts and PSF count within indicated radius.
d) Fraction of counts in the extended component relative to the PSF counts.

In the 3-7 keV band, the expected number of background counts in the three apertures ranges from ≤1 to ~13, and is much less in the narrower bands, especially at energies >4 keV. Table 2 lists the results for incremental 1-keV-wide energy bands, above 3 keV. The difference in our results from using the pure mirror PSF may affect the Δ(1".5) estimates, but in any case is within the errors of Table 2. In conclusion, we find a trend of more extended components at the lower energies. This trend is consistent with the finding of Paper II for the full extent of the diffuse emission. However, the overall fraction of the emission in the extended component is larger in these inner regions.

## 8. Discussion

In Elliptical and cD galaxies in clusters it is well established that radio jets and lobes produced by the central supermassive black hole carve cavities in the hot ISM and transfer enough energy to halt strong cooling flows (e.g., Russell et al 2013). In contrast, for radio-faint Seyfert galaxies the warm absorber outflows rarely transfer more that ~1% of energy and momentum to the interstellar medium (Wang et al 2011c, Crenshaw & Kraemer 2012). Instead, molecular outflows typically dominate AGN feedback, ejecting up to 10% of $L_{bol}$(AGN) (Fiore et al. 2017). The mechanism coupling the AGN to the molecular ISM is complex, and includes photoionization, and collisional ionization (shocks) from the interaction with small radio jets. This process may also generate large-scale winds. Multiphase hot/cold interactions are important to understanding feedback (Mukherjee et al. 2016).

The deep high-resolution *Chandra* images of the CTAGN ESO 428-G014 provide a detailed view of the inner high surface brightness regions that can contribute to the understanding of AGN-galaxy interaction. Our analysis has revealed sub-arcsecond scale features in the emission-line dominated soft band (<3 keV) that we have compared in detail with the high-resolution optical emission line and radio continuum maps (Falcke et al 1996, 1998). The implications of these results for the characterization of the physical status of the emitting excited medium (Section 8.1), and our understanding of jet-ISM interaction (Section 8.2) are discussed below.

We have also revisited the relative importance of unresolved AGN emission versus diffuse hard emission in the energy band >3 keV, which is dominated by the hard continuum emission and the 6.4 Fe Kα line. In this paper, we analyze the innermost resolved regions (<1".5 ~ 170 pc) that were not considered in Paper II. These results both provide information on the circumnuclear ISM and set caution on modeling of the AGN (Section 8.3).

### 8.1 Photoionization and collisional ionization

The spectral analysis of Paper II has shown that the soft (<3 keV) X-ray emission of ESO 428-G014 is complex, requiring a mix of multiple photoionization and thermal components to account for the observed X-ray emission lines. This analysis also shows that the inferred density of the X-ray emitting plasma is a few atoms per cubic centimeter, consistent with what may be expected from a typical spiral disk ISM. These results suggest that the emission is likely to be due both to ionization of ISM clouds by AGN photons, and to shocks. The interaction of the radio jet (Falcke et al 1996, 1998) with ISM clouds may be a cause of collisional ionization. Similar dual ionization mechanisms have been reported in our studies of NGC 4151 (Wang et al 2011b) and Mkn 573 (Paggi et al 2012).

The overall good correspondence between Hα, [O III] and X-ray maps described in this paper for ESO 428-G014 is consistent with the photoionization scenario. However, the existing *HST* archival data do not provide an exposure of the continuum, and therefore we cannot calculate [O III]/X-ray ratios, to compare with other Seyfert galaxies and individual emitting clouds, as in e.g. NGC 4151 (Bianchi et al 2006; Wang et al 2011a, c). In NGC 4151, in particular, the majority of clouds that could be measured with *Chandra* and *HST* are consistent with the same [O III]/X-ray ratio, i.e. a constant ionization parameter, suggestive of an expanding wind (Wang et al 2011c). Instead, a localized increase of the X-ray emission (smaller [O III]/X-ray ratios) is found in the areas of ISM-radio jet interaction.

The X-ray emission line analysis of Section 3.1 (shown in Fig. 14) directly demonstrates how the spectral complexity suggested by model fitting of the bulk X-ray spectrum (Paper II) may be related to cloud-to-cloud differences in ESO 428-G014. In the tail (see Fig. 7) there is a prominent knot of O VII emission to the SE. The nucleus is prominent in both O VII and Si XIII, while the NW knot in the head is mostly prominent in the Ne IX and X lines, and somewhat also in Si XIII.

The prominent OVII SE knot in the tail (blue in Fig. 14) is consistent with a lower ionization parameter and lower $N_H$ than the nucleus (red) and the NW knot (yellow). For a static ISM, this may be understood given its greater distance from the nucleus. However, this conclusion would contradict the results of NGC 4151, of a wind expanding in the ionization bicone (Wang et al 2011c). This expanding wind may be present in ESO 428-G014, where a biconical outflow with velocity of ~1400 km s$^{-1}$ is suggested by the optical line spectra (Wilson and Baldwin 1989). Moreover, the NW knots resides in a relatively lower excitation region than the SE knot, based on the [OIII] / Hα contours of Falcke et al (1996).

It is however possible that the excess of Ne IX+X emission in the NW knot is at least in part due to collisional excitation. Simulation of thermal models (APEC, for the range of parameters suggested by Paper II), show a marked increase of this line emission, relative to O VII and Si XIII for kT~1 keV. Wang et al (2011b), Paggi et al (2012) and Maksym et al (2018) show increased Ne line emission in correspondence with the termination of radio jets in NGC4151, Mkn 573 and NGC 3993, respectively. Moreover, in ESO 428-G014, Falcke et al (1998) suggest that shock emission may be linked to the Hα emission in the NW area. We should mention that the Ne lines were identified on the basis of the spectral fit to the ACIS data (Paper II). However, we cannot exclude some contamination with Fe XIX lines. For example, this is reported by Koss et al (2015) in the inner regions of NGC 3993, observed with the High Energy Transmission Gratings (HETG). Similar data for ESO 428-G014 are not available.

The nucleus has line ratios outside of the model grid, but in the direction of increasing $N_H$ (see the right panel of Fig. 14), which is consistent with a local higher gaseous density. This is not surprising, given the high nuclear obscuration of CT AGNs (see also Fig. 6).

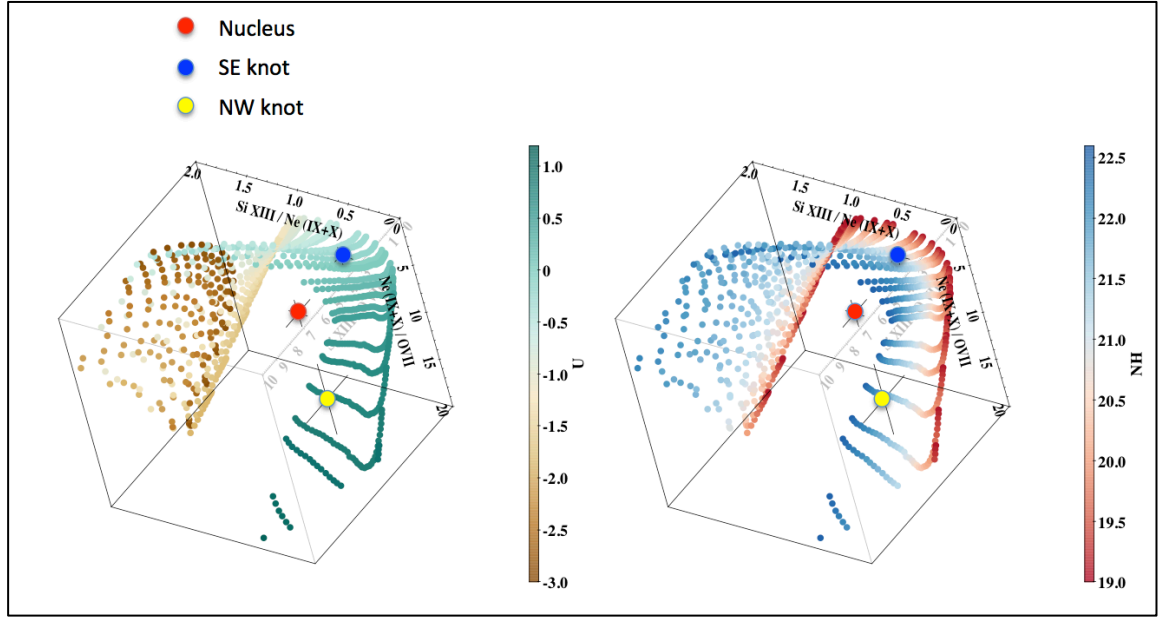

Fig. 14 – Line ratios of three emission areas from the images (see Section 3.1, Fig. 7; Nucleus, red; SE knot, blue; NW knot, yellow) plotted on photoionization model predictions for a range of log U (color scale, left) and log $N_H$ (color scale, right), as indicated by the respective color scales. The three axes of the cube are: Si XIII/Ne (IX+X), Ne (IX+X)/O VII, and in the background Si XIII/O VII. The measured ratios are plotted with their 1σ error bars. In the case of the SE knot the errors are commensurable with the size of the point.

## 8.2 Jet – ISM interaction

The overall soft emission (0.3-3.0 keV) in the circumnuclear 'head' (Section 3) has a complex morphology, which roughly echoes that of the optical line emission (Hα and [O III]), and is generally spatially correlated with the jet radio continuum emission (Section 5). These inter-related features of the radio jet, the warm and the hot ISM are consistent with the predictions of recent relativistic hydrodynamics simulations by Mukherjee et al (2018). Tailored to the radio-loud disk galaxy IC 5063, these simulations reproduce the observed clumpy morphology of the cool (CO) and warm ([OIII]) ISM in the region of jet interaction and show that both warm and hot (X-ray) emission are expected in the jet-cold disk interaction region. The X-ray morphology of the disk will depend on jet power and age. Although the jet power in ESO 428-G014 may be 1/10 to 1/100 of that in IC 5063 (based on the respective radio fluxes and distances of these galaxies; Morganti et al 2015, Leipski et al 2006, Condon et al 1998), the correspondence between radio, optical emission line, and X-ray features we have found are intriguingly similar to those predicted by Mukherjee et al (2018).

These simulations also predict that $\sim 10^7$ K gas would be forming at later stages of the interaction. This gas can be found both around the cooler gas clumps, and being expelled outside of the disk in a large-scale filamentary wind (Fig. 4 of Mukherjee et al 2018). While the spectral properties of the extended X-ray emission of ESO 428-G014 are mostly consistent with photoionization, we cannot exclude the presence of thermal emission in the largest spatial scales (> 1 kpc), given the low signal to noise of this emission, which precludes a spectral characterization. We find in Paper II a low-temperature (<1.5 keV) ISM component. We also find some suggestion of filamentary protrusions in the low surface brightness regions (see Figs. 1, 2, 10).

In Paper II and in our previous work on NGC 4151 and Mkn 573 (Wang et al 2011c, Paggi et al 2012), we found significant emission in the direction perpendicular to the photoionization cone, which should be obscured by the torus in the standard AGN model. We speculated that this emission may indicate AGN torus porosity. The jet-ISM interaction models (see e.g., Mukherjee at al 2018, Wagner and Bicknell 2011) also predict the formation of a very hot ($\sim 10^{8-9}$ K cocoon around the nucleus). It would be interesting to investigate if the predicted emissivity of this cocoon could contribute to (or explain) the X-ray excess we see in the cross-cone direction of the AGNs we have studied with *Chandra*.

Our high-resolution images show an enhancement of the X-ray emission in the area of second 'crossing' of the H$\alpha$ and [OIII] strands (Figs. 9, 10). The EMC2 restoration of the X-ray image (Fig. 8) suggests a double loop structure emanating from this bright spot that may be connected with the morphology of the H$\alpha$ strands in this area. This is also the region of enhanced Ne line emission, possibly connected with thermal emission from collisional shocks (Section 8.1, Fig. 14). These results would support our suggested association of the NW bright radio jet with this region (Fig. 12, Left). As mentioned in Section 8.1, shocked emission at the termination of small radio jets has been found in at least other three nearby Seyfert galaxies with *Chandra*.

**8.3 Extended versus point-like hard emission in the inner regions**

The morphology of the emission in the circumnuclear region (the head, see Section 3) at energies > 3.0 keV (i.e. spectral hard continuum and Fe K$\alpha$ line, Papers I and II), shows the emergence of a prominent point source in the hard continuum at energies > 4 keV (Fig. 7). However, even in this inner region, a large fraction of the emission is from a more extended component (Section 7, Table 2).

The presence of hard diffuse emission is consistent with the results of Papers I and II, where extended hard continuum and Fe K$\alpha$ line components were reported from significantly larger, kpc-size radii.

In Paper II we reported that the diffuse emission detected at energies > 3 keV in a circumnuclear annulus of 1.''5 – 8'' radii (~170 – 900 pc) amounts to ~30% of the total hard emission of ESO 428-G014. From the present analysis, it is clear that the extended emission is even more important in the central circum-nuclear region (r < 170 pc), amounting to ~70 - 30% of the contribution of the nuclear point source in that area (or ~40 - 25% of the total counts in the region). Within a 5" radius (~560 pc), the contribution from the extended emission relative to the expected point source emission increases; it overcomes that from the nuclear point source in the 3 - 4 keV band. Table 2 shows that within the central ~170pc and ~560pc radius regions the diffuse component contributes relatively less at the higher energies. This may reflect the relative composition of cloud densities in the ISM.

This extended emission is likely to be from scattering of nuclear photons by dense molecular clouds in the inner disk of ESO 428-G014 (Paper II), and it is detectable only with the sub-arcsecond imaging capability of *Chandra*. Ignoring its presence will bias the spectral modeling of the AGN torus as this emission will be wrongly attributed to the torus, and hence will lead to an overestimate of the derived covering factor and may affect the derived geometry (e.g., Bauer et al. 2015; Baloković et al. 2018) with spectra from X-ray telescopes with inferior angular resolution, such as *NuSTAR* and *XMM-Newton*.

## 9. Summary and Conclusions

We have analyzed the deep *Chandra* observation (~155 ks) of the CT AGN ESO 428-G014 to study in details the morphology of the diffuse X-ray emission in the inner ~500 pc radius region, as function of energy. In summary, we find that:

1) The central 500 pc region of ESO 428-G014 contains the NW-SE high surface brightness X-ray emission (Paper II). This region has a head-tail morphology, where the head comprises the 1" (112 pc) radius circumnuclear region, and the tail extends to the SE for ~4". The tail is more prominent at the lower energies, below 3.0 keV, consistent with the general energy-dependence of the extended emission discussed in Paper II.

2) Taking the ratio of images in different energy bands, we find localized relatively hard emission regions. One such region is a ~200pc long feature extending NE to SW across the nucleus, perpendicular to main axis of the high surface brightness X-ray emission. The physical size and orientation of this feature is intriguingly similar to that of the obscuring torus of NGC 4945 (Marinucci et al 2012, 2017), however in our case we do not see excess emission directly in the hard-band (3.0-6.0 keV). We also find relatively harder emission (1.5 -3.0 keV) to the N of the 'head' and S of the 'tail'. This features echo the morphology of the J-K colors image (Riffel et al 2006), suggesting a connection with absorption of the softer photons in these areas.

3) Narrow-band images in the most prominent soft X-ray lines (O VII - 0.40-0.62 keV; Ne IX + Ne X - 0.9-1.2 keV; and Si XIII - 1.65-2.0 keV; see Paper II), show localized differences. Comparing the observed line ratios with a set of photoionization models, we find statistically significant differences in the ionization state of the emitting plasma of these regions (Fig. 14). This is consistent with the presence of several spectral components in the overall emission, suggested by the spectral fits of Paper II. The line ratios for the nuclear region are consistent with the larger absorption $N_H$ expected in this CT AGN.

4) Using the positions of the hard X-ray nucleus (4-6 keV) and the optical nucleus (Falcke et al 1996) to match the images, we find a good correspondence between optical line (H$\alpha$ and [OIII]; Falcke et al 1996) and soft (<3 keV) X-ray features. The areas of stronger X-ray emission in the head are spatially coincident with the 'crossing' in the H$\alpha$ strands. This is also the area of most intense localized [OIII] emission, and based on the X-ray emission line ratios, could include shocked thermal X-ray emission. The bubble-like frothy features in the head suggested by the EMC2 restored X-ray image are consistent with the position of the NW loop in the optical line emission. Enhancements in surface brightness in the tail to the SE are spatially consistent in H$\alpha$, [OIII], and 0.3-3.0 keV bands.

5) On the larger $\sim$ 1kpc scale, the overall agreement between optical line contours and X-ray brightness levels is remarkable, except for the SE corner, where the [OIII] emission stretches further out (as noted by Falcke et al 1996, see Fig. 10 bottom left, showing the [OIII]/H$\alpha$ contours). The 6 cm radio jet also follows this region. The upturn of the 6 cm radio jet (Fig. 10 bottom right) occurs just past the region of intense H$\alpha$ and X-ray emission shown in Fig. 9, in proximity of the upturn in the Si XIII band emission (Fig. 7).

6) While the relative astrometry of the optical line and X-ray images is unequivocal, there is some uncertainty on the comparison with the high-resolution 2 cm VLA radio image. Falcke et al (1998), based on the H$\alpha$ and radio morphology, associated the nucleus with the SE bright radio knot. However, in the central $\sim$100pc, the comparison between hard X-ray continuum (5-6 keV) and radio images suggests (but does not prove) that the central fainter radio knot may be alternatively associated with the nucleus, as identified by the hard point-like nuclear X-ray source. In this case, the NW bright radio knot would coincide with the higher surface brightness region in the soft X-ray image, which is also coincident with the second crossing of the H$\alpha$ strands, and with the area of possibly shocked X-ray emission (based on excess Ne IX + Ne X line emission). The SE bright radio knot would be coincident with the region of rapid gradient of X-ray brightness. Based on our previous studies of nearby Seyfert galaxies, the association of the NW radio knot with an enhancement in the Ne IX and X emission would favor the association of this region with the termination of the radio jet (Wang et al 2011b, Paggi et al 2012, and Maksym et al 2018).

7) At all energies >3 keV there is a large contribution of extended emission in the central 1.5" ($\sim$170 pc) radius central circumnuclear region, amounting to $\sim$70-30% of the contribution of a point source in that area (or $\sim$40-25% of the total counts in the region). Within a 5" radius ($\sim$560 pc), the contribution from extended emission increases at lower energies, and it overcomes that from a nuclear point source in the 3-4 keV band. This extended emission is likely to be from scattering of nuclear photons by dense molecular clouds in the inner disk of ESO 428-G014 (Paper II). It may adversely bias the torus modeling of spectra from X-ray telescopes with inferior angular resolution *than Chandra*, such as *NuSTAR* and *XMM-Newton*.

8) The overall similarity of the morphology of the radio jet, the warm and the hot ISM are consistent with the predictions of recent relativistic hydrodynamics simulations by Mukherjee et al (2018) of the interaction of radio jets with a dense molecular disk. These simulations also predict a hot cocoon enveloping the interaction region, offering a possible alternative explanation to the presence of X-ray emission perpendicular from the bicone. In Paper II we suggested that a porous circum-nuclear torus may cause this emission.

This work demonstrates how spectral-spatial mapping of AGNs in the X-ray band provides a new and unique view of the AGN-host galaxy interaction that complements the high-resolution views achievable in the optical and radio bands. The ultimate sub-arcsecond spatial resolution of *Chandra* is essential for pursuing these studies, as is the high-count statistics that can only be obtained with long *Chandra* exposures. These results make the case for a future large-collecting-area X-ray observatory that preserves the angular resolution of Chandra, such as *Lynx*.

We thank Heino Falcke for a conversation on the nuclear astrometry. We thank John Raymond, Aneta Siemiginowska, and Marta Volonteri for comments on the manuscript. We retrieved data from the NASA-IPAC Extragalactic Database (NED) and the Chandra Data Archive. For the data analysis, we used the CIAO toolbox, Sherpa, and DS9, developed by the Chandra X-ray Center (CXC); and XSPEC, developed by the HEASARC at NASA-GSFC. This work was partially supported by the Chandra Guest Observer program grant GO5-16090X (PI: Fabbiano) and NASA contract NAS8- 03060 (CXC). The work of J.W. was supported by the National Key R&D Program of China (2016YFA0400702) and the National Science Foundation of China (11473021, 11522323).